\documentclass[aps,pra,twocolumn]{revtex4}%
\usepackage{graphics}
\usepackage{amsmath}
\usepackage{graphicx}
\usepackage{amsfonts}
\usepackage{amssymb}
\usepackage{xcolor}%
\begin{document}

\title{Towards practical security of continuous-variable quantum key distribution}
\author{Cosmo Lupo}
\affiliation{Department of Physics \& Astronomy, University of Sheffield, UK}

\begin{abstract}
Rigorous mathematical proofs of the security of continuous-variable quantum key distribution (CV QKD) have been obtained recently.
Unfortunately, these security proofs rely on assumptions that are hardly met in experimental practice. 
%
Here I investigate these issues in detail, and discuss experimentally-friendly workarounds to assess the security of CV QKD.
The aim of this paper is to show that there are hidden and unsolved issues and to indicate possible partial solutions. To provide a complete and rigorous mathematical security proof is out of the scope of this contribution.
\end{abstract}

\maketitle

\section{Introduction}\label{Sec:intro}

Quantum key distribution (QKD) is a family of experimental methods that exploit quantum optics to realize the task of secret key expansion \cite{Scarani2009,Diamanti,Bacco}.
Early works on QKD were defined within a discrete-variable (DV) architecture, where information is encoded in discrete degrees of freedom (e.g., polarization, phase delay, etc.)
of a single photon or weak coherent pulse.
This requires single-photon detectors at the measurement stage. 
Later works introduced continuous-variable (CV) architectures, where information is encoded in a continuous way in the quadratures of the electromagnetic field, and homodyne or heterodyne detection can be utilized at the measurement stage \cite{EReview}. 
The fact that CV QKD utilizes homodyne or heterodyne detection, which are mature detection techniques
routinely deployed in optical communications, is considered to be the strategic advantage of CV QKD over DV QKD \cite{Diamanti}.

Obtaining security proofs for CV QKD protocols is known to be a particularly challenging task. 
The difficulties are not conceptual, but technical, and stem from the fact that the underlying Hilbert space has infinite dimensions.
The first complete and rigorous security proofs were obtained by
Furrer {\it et al.}\ in Refs.\ \cite{Furrer,FurrerRR}
and by Leverrier in Refs.\ \cite{Lev1,Lev2}.
Refs.\ \cite{Furrer,FurrerRR} considered a protocol where Alice prepares squeezed states and Bob measures by homodyne detection,
and the security proof was obtained by applying entropic uncertainty relations.
Refs.\ \cite{Lev1,Lev2} (see also Ref.\ \cite{Lev2way}) considered a protocol where Alice prepares coherent states whose amplitudes are sampled from a Gaussian distribution, and Bob measures by heterodyne detection \cite{no-switching}.

This paper focuses on a protocol with coherent states in input and heterodyne detection in output.
I will discuss the discrepancies between the mathematical model used in the security proof of Ref.\ \cite{Lev1} and the actual experimental practice.
In particular, I will focus on two main experimental limitations:
\begin{enumerate}
    \item It is practically impossible for the sender Alice to sample from a continuous Gaussian distribution. In any experimental implementation of the protocol she will sample from a discrete and bounded distribution of coherent state amplitudes.
    \item Experimental heterodyne detection differs from the mathematical model used in the security proofs. In particular, the range of ideal heterodyne detection is unbounded, whereas its experimental implementations have necessarily a finite range.
\end{enumerate}

The reason to focus on these two non-idealities is two fold: 
1) they play a central conceptual role in the proof strategy of Refs.\ \cite{Lev1,Lev2};
2) they represent a serious challenges to experimental demonstrations of secure CV QKD.

The first issue about state preparation was addressed by Jouguet {\it et al.}\ in Ref.\ \cite{Jouguet}. Here I show that their approach is not experimentally feasible.
More recently, this issue has been addressed by Kaur, Guha, and Wilde in Ref.\ \cite{Wilde}, who provided a solution that depends on three free parameters that cannot be estimated from experimental data. 
The approach I develop here, which is conceptually similar, reduces the number of free parameters from three to two. 
Furthermore, I present a physical argument that may allow us to get rid of the remaining two free parameters.
To the best of my knowledge, the second issue about non-ideal heterodyne detection has never been explicitly addressed in literature.


The aim of this work is {\it not} to present a complete and rigorous mathematical security proof, but to spell out the limitations of known security proofs and suggest an approach that may, in some part, solve them.
In particular, I do not discuss parameter estimation here. This will allow me to focus on the two physical issues outlined above. 
Parameter estimation has been extensively discussed in other works, for example Ref.\ \cite{Lev1} and references therein.
Finally, this work focuses on assessing the security against collective attacks.
More work is needed to extend this approach to general, coherent attacks. Following the results of Ref.\ \cite{Lev2}, we may expect that extension to coherent attacks comes with significant overheads.

The paper develops as follows.
Section \ref{Sec:ideal} reviews the ideal protocol of Ref.\ \cite{no-switching}.
Section \ref{Sec:deviations} discusses in detail the two main discrepancies between ideal CV QKD and its experimental realizations.
Section \ref{Sec:practical} presents a particular realization of CV QKD protocol, of which I present a security analysis.
Section \ref{Sec:toolbox} reviews the main theoretical tools utilized.
Section \ref{Sec:proof} presents the security analysis.
Section \ref{Sec:coherent} briefly discusses a potential extension to coherent attacks.
Examples are discussed in Section \ref{Sec:example}.
A summary of the results is given in Section \ref{Sec:summary}.
Section \ref{Sec:conclusions} is for conclusions.

\section{Description of the ideal protocol}\label{Sec:ideal}

This section reviews the main steps of the CV QKD protocol as it would be implemented ideally, i.e., using ideal experimental devices. 
I will do this with the help of the familiar fictional characters 
Alice (the authenticated sender of quantum signals), Bob (the authenticated receiver), and Eve (the eavesdropper).
The ideal protocol is essentially the one first proposed by Weedbrook {\it et al.}\ in Ref.\ \cite{no-switching}:

\begin{enumerate}

\item {\it Quantum state preparation.}
The sender Alice samples random numbers $q_A$, $p_A$ from a Gaussian distribution with zero mean and variance $N$.
Alice then prepares a coherent state $|\alpha\rangle$ 
with amplitude 
$\alpha = ( q_A + i p_A )/\sqrt{2}$.
Physically, $N$ represents the mean photon number in the input states.

The ensemble of input coherent states prepared by Alice on system $A$
is represented by a bipartite state,
\begin{align}\label{idealXA}
\sigma_{XA}^0 = \int d^2\alpha P^0(\alpha) | u_{\alpha} \rangle \langle u_{\alpha} | \otimes  | \alpha \rangle \langle \alpha |  \, ,
\end{align}
where 
\begin{align}\label{Gaussian}
P^0(\alpha) = \frac{1}{\pi N} \, e^{-|\alpha|^2/N} \, ,
\end{align}
$d^2\alpha = \frac{1}{2} dq_A dp_A$,
and $| u_{\alpha} \rangle$ is a dummy quantum state that 
carries the value of $\alpha$.
The latter is represented as the random variable $X$.
 
\item {\it Quantum communication.} 
Alice sends the coherent state to Bob through an
untrusted quantum communication channel that may be controlled by Eve.

The action of the quantum channel on system $A$ is described as
an isometry transformation $U_{A \to BE}$ that broadcasts
quantum information to Bob ($B$) and Eve ($E$).
This is represented by the tripartite state,
\begin{align}\label{idealXBE}
\sigma_{XBE}^0 = \int d^2\alpha P^0(\alpha) | u_{\alpha} \rangle \langle u_{\alpha} | \otimes \psi_{BE}^0(\alpha) \, ,
\end{align}
where $|\psi_{BE}^0(\alpha)\rangle = U_{A \to BE} |\alpha\rangle$ is the joint state of Bob and Eve for given $\alpha$, and $\psi_{BE}^0(\alpha)$ is a short hand notation for $|\psi_{BE}^0(\alpha)\rangle \langle \psi_{BE}^0(\alpha)|$.

\item {\it Measurement.}
Bob measures the received signal by ideal heterodyne detection.
The measurement outcome is a pair of real numbers, denoted as $q_B$, $p_B$, that can be represented as a complex number
$\beta = ( q_B + i p_B )/\sqrt{2}$.

The correlations between $\alpha$, $\beta$ an Eve's quantum side information are described by the state
\begin{align}\label{idealXYE}
& \sigma_{XYE}^0 = \nonumber \\
& \int d^2\alpha d^2\beta P^0(\alpha,\beta) | u_{\alpha} \rangle \langle u_{\alpha} | \otimes | v_{\beta} \rangle \langle v_{\beta} | \otimes \psi_{E}^0(\alpha,\beta) \, ,
\end{align}
where $| v_\beta \rangle $ is a dummy quantum state
that carries the value of $\beta$, which is represented
by the random variable $Y$,
$\psi_{E}^0(\alpha,\beta)$ is Eve's state conditioned on $\alpha$ and $\beta$, 
and $P^0(\alpha,\beta) = P^0(\alpha) P^0(\beta|\alpha)$ is the joint probability.

\end{enumerate}

A symmetrization step could be added to the protocol, see Ref.\ \cite{Lev1}, to simplify the parameter estimation routine.
For the sake of simplicity I do not consider this step here, also because a discussion of parameter estimation is out of the scope of this work.

A quantity of particular interest is the covariance matrix of the quadratures:
\begin{align}\label{CM0}
\gamma_{jk}^0 := \langle Q_j Q_k \rangle_{P^0} - \langle Q_j \rangle \langle Q_k \rangle_{P^0} \, ,
\end{align}
where $Q_j,Q_k \in \{ q_A, p_A, q_B, p_B \}$, and
\begin{align}
\langle F(Q_j,Q_k) \rangle_{P^0} := \int d^2\alpha d^2\beta P^0(\alpha,\beta) F(Q_j,Q_k) \, ,
\end{align} 
for any function $F$.

The above three steps need to be repeated $n$ times.
After that, Alice and Bob post-process their local raw data to extract a secret shared key of $\ell$ bits. 

Given the physical parameters that characterize the protocol, including the noise and loss associated with the communication channel, the value of $n$ is one of the factors that determine how many secret bits can be generated.
For typical noise and loss values, $n$ could be as large as $10^8 - 10^{12}$ \cite{Jouguet,Jouguet2,Jouguet3,Jouguet4,Lev1}. 
This strongly depends on loss, noise, and on the required standard of security.

The classical post-processing includes the routines of
parameter estimation, error reconciliation,
and privacy amplification.
Here I assume (without loss of generality) that Alice reconciles her raw data with Bob (reverse reconciliation).
In order to do this efficiently, Alice and Bob need to apply an Analog to Digital Converter (ADC)
to discretize the variables $X$ and $Y$.
We denote as $\bar X$ and $\bar Y$ the discretized variables, their values will determine the raw keys of Alice and Bob, respectively.

The ADC is characterized by its range $R$ and number of output bits.
For example, the ADC on Bob's side is defined by a set of $d$ non overlapping intervals,
with $d$ equal to the cardinality $|\bar Y|$ of $\bar Y$, 
\begin{align}
I_j = ( - R + (j-1)\delta , - R + j\delta ] \, , 
\end{align}
for $j=1,\dots, d-2$, and 
\begin{align}
I_0 & = ( - \infty , - R  ] \, , \\
I_{d-1} & = ( - R + (d-2)\delta , + \infty) \, ,
\end{align}
with $\delta = 2R/(d-2)$.
To each pair of intervals we associate a unique amplitude value $\beta_{jk} = ( q_{Bj} + ip_{Bk} )/\sqrt{2}$, 
where $q_{Bj} = - R - \delta/2 + j\delta$ and $p_{Bk} = - R -\delta/2 + k\delta$.

Finally, we obtain a description of the state of Bob and Eve after the ADC
\begin{align}\label{idealbarYE}
& \sigma_{\bar YE}^0 = \nonumber \\
& \sum_{jk} | v_{\beta_{jk}} \rangle \langle v_{\beta_{jk}} | \otimes 
\int d^2\alpha \int_{I_j \times I_k} d^2\beta 
P^0(\alpha,\beta)  
\psi_{E}^0(\alpha,\beta) \, ,
\end{align}
where $I_j \times I_k$ denotes the set of values of $\beta$ such that $q_{B} \in I_j$ and $p_{B} \in I_k$.

\section{Deviations from the ideal protocol} \label{Sec:deviations}


This section presents a (not exhaustive) list of the discrepancies between the ideal protocol and its physical implementations.
This will focus on how state preparation and measurement are modelled in the security proof of CV QKD.

\subsection{Deviation from ideal quantum state preparation}

This accounts for the fact that it is not physically possible to sample coherent states with a Gaussian distribution of their amplitudes. This is because any physical device operates within a finite range and resolution.

In any physical realization of the protocol, Alice samples the coherent state amplitudes from a discrete and bounded distribution.
A known way to assess the security of the protocol with discrete amplitude modulation is to consider the statistical distance between the average input states of the ideal and practical protocols \cite{Jouguet}.
In the ideal protocol, the average state that Alice sends to Bob is
\begin{align}
\sigma_A^0 = \frac{1}{\pi N} \int d^2\alpha \, e^{-\frac{|\alpha|^2}{N}}|\alpha\rangle \langle \alpha| \, .
\end{align}
The state $\sigma_A^0$ is in fact a thermal state with $N$ mean photons.

In practice, Alice draws the coherent state amplitudes from some discrete and bounded ensemble, $\{ p(j), \alpha_j \}_{j=1,\dots,\nu}$,
where $p(j)$ is the probability of the complex amplitude $\alpha_j$.
Therefore, the average state sent to Bob reads:
\begin{align}
\sigma_A = \sum_{j=1}^\nu p(j) |\alpha_j\rangle \langle \alpha_j| \, .
\end{align}

To compare the ideal with the experimental state preparation step, one considers the trace distance \cite{Jouguet}
\begin{align}\label{Eq1:td}
D(\sigma_A, \sigma_A^0) 
= \frac{1}{2} \mathrm{Tr}\left| \sigma_A - \sigma_A^0 \right| \,.
\end{align}

Recall that the trace distance quantifies the probability of successfully discriminating the two states (see, e.g., Ref.~\cite{MDI_2018_1}). 
A bound on the trace distance of the form $D(\sigma_A, \sigma_A^0) \leq \epsilon^{(1)}$ implies that any attempt to distinguish $\sigma_A$ from $\sigma_A^0$ succeeds with probability no larger than $\epsilon^{(1)}$.

Because the protocol requires the preparation of $n$ signals, we should not consider the quantity in Eq.\ (\ref{Eq1:td}), but its $n$-fold version,
\begin{align}\label{Eqn:td}
D(\sigma_A^{\otimes n}, {\sigma_A^0}^{\otimes n}) \, ,
\end{align}
which is defined on $n$ identical copies of $\sigma_A$ and $\sigma_A^0$.
This is related to the single-copy trace distance 
through the inequality 
\begin{align}
D(\sigma_A^{\otimes n}, {\sigma_A^0}^{\otimes n}) \leq n D(\sigma_A, \sigma_A^0) \, .
\end{align}
Therefore, the practical protocol is indistinguishable from the ideal one up to a
probability smaller than $\epsilon^{(n)} = n \epsilon^{(1)}$.
Note that this failure probability grows linearly with $n$.

We want the failure probability to be sufficiently small.
For example, some authors put the overall security failure probability 
in a range between $10^{-10}$ \cite{Jouguet,Jouguet2,Jouguet3}
and $10^{-20}$ \cite{Lev1}. 
Putting this together with the fact that $n$ may be in the range of $10^8 - 10^{12}$,
we obtain that $\epsilon^{(1)}$ needs to lay somewhere between
$10^{-18}$ and $10^{-32}$. 

Obviously, achieving this target would require a level of experimental control that is hardly seen in laboratory practice.
This shows that, despite recent mathematical results, the question still remains open: {\it How CV QKD can be made provably secure in practical experimental realizations?}


\subsection{Deviation from ideal heterodyne detection}

This accounts for the fact that any physical device that implements Bob's measurement has finite range, whereas the output of ideal heterodyne detection is unbounded.


This is a most important issue because it affects the cornerstone of the security proofs for CV QKD protocol, i.e., the optimality of Gaussian attacks \cite{GarciaPatron,Navascues}.
This important result, which will be reviewed in the Section \ref{Sec:toolbox}, only holds for ideal heterodyne detection.
In fact, the property of optimality of Gaussian attacks follows from the symmetry of ideal heterodyne detection. This symmetry is broken when ideal heterodyne is replaced with non-ideal heterodyne having a finite range.

It is not clear how it could be extended or adapted to finite-range non-ideal heterodyne.

In other words, the celebrated extremality of Gaussian attacks holds under the assumption of ideal homodyne or heterodyne detection. In reality, the experimental realization of these detection techniques is imperfect, as they have a bounded range. {\it The conclusions then is that, strictly speaking, the property of extremality of Gaussian attacks cannot be applied to any physical realization of CV QKD.}

\section{Description of a practical protocol}\label{Sec:practical}

We consider a specific experimental scheme for quantum state preparation, where Alice prepares coherent states drawn from a finite set. See Table \ref{Table:protocol} for a summary of the notation used.

Consider a set of $d$ non-overlapping intervals:
\begin{align}
J_j = ( - R_A + j\delta_A , - R_A + (j+1)\delta_A ] \, , 
\end{align}
for $j=0,\dots, d-1$, 
with $\delta_A = 2R_A/d$.
To each pair of these intervals we associate a unique complex
number
$\alpha_{jk} = ( q_{Aj} + ip_{Ak} )/\sqrt{2}$, 
where $q_{Aj} = - R_A + j\delta_A + \delta_A/2$ 
and $p_{Ak} = - R_A + k\delta_A + \delta_A/2$.

The state preparation routine is then defined as follows.
First Alice draws a complex value $\alpha = (q_A + ip_A)/\sqrt{2}$
from a (sub-normalized) probability distribution $P(\alpha)$ on the domain 
$\mathcal{R}_A = [-R_A,R_A] \times [-R_A,R_A]$, i.e., $P(\alpha)$
is non-zero only if $q_A \in [-R_A,R_A]$ and $p_A \in [-R_A,R_A]$.
Then, she prepares the coherent state 
$|\alpha_{jk}\rangle$ if $q_A \in J_j$ and $p_A \in J_k$.

The ensemble prepared in this way is described by the bipartite state
\begin{align}\label{realXA}
\sigma_{XA} 
= \sum_{j,k=0}^{d-1} \int_{J_j \times J_k} d^2\alpha P(\alpha) |u_{\alpha} \rangle \langle u_\alpha | \otimes |\alpha_{jk} \rangle \langle \alpha_{jk} | \, .
\end{align}
We shall compare this with its counterpart $\sigma_{XA}^0$ for the ideal protocol in Eq.\ (\ref{idealXA}), using
for example, the trace distance 
\begin{align}\label{Error_p}
\epsilon_\mathrm{p} := D(\sigma_{XA}, \sigma_{XA}^0) \, .
\end{align}


We also define the states $\sigma_{XBE}$ and $\sigma_{XYE}$ for the practical protocol in the same way as we have done for the ideal protocol. 
We have
\begin{align}
\sigma_{XBE} 
= \sum_{j,k=0}^{d-1} \int_{J_j \times J_k} 
d^2\alpha P(\alpha) 
|u_{\alpha} \rangle \langle u_\alpha | 
\otimes \psi_{BE}(\alpha_{jk}) \, ,
\end{align}
and
\begin{align}
\sigma_{XYE} & = \sum_{j,k=0}^{d-1} \int_{J_j \times J_k} 
d^2\alpha 
\int d^2\beta
P(\alpha,\beta) \nonumber \\
& \hspace{2cm}
|u_{\alpha} \rangle \langle u_\alpha | 
\otimes 
|v_{\beta} \rangle \langle v_\beta | 
\otimes 
\psi_{E}(\alpha_{jk},\beta) \, ,
\end{align}
where $P(\alpha,\beta)$ is the joint probability of Alice and Bob. 

It follows from the monotonicity of the
trace distance under completely positive maps that
\begin{align}
D(\sigma_{XBE}, \sigma_{XBE}^0) & \leq \epsilon_\mathrm{p} \, , \\
D(\sigma_{XYE}, \sigma_{XYE}^0) & \leq \epsilon_\mathrm{p} \, ,
\end{align} 
and
\begin{align}\label{Pdistance}
D(\sigma_{XY}, \sigma_{XY}^0) = \frac{1}{2} \int d^2\alpha d^2\beta 
\left| P(\alpha,\beta) - P^0(\alpha,\beta) \right|
\leq \epsilon_\mathrm{p} \, .
\end{align} 
This latter bound will be useful for the estimation of the cross-diagonal terms of the covariance matrix in Section \ref{Sec:issue2}.

We define the covariance matrix for the practical protocol, 
\begin{align}\label{CM}
\gamma_{jk} := \langle Q_j Q_k \rangle_{P} - \langle Q_j \rangle \langle Q_k \rangle_{P} \, ,
\end{align}
with 
\begin{align}
\langle F(Q_j,Q_k) \rangle_{P} = \int d^2\alpha d^2\beta P(\alpha,\beta) F(Q_j,Q_k) \, .
\end{align}
Comparing this with the covariance matrix $\gamma^0$
in Eq.\ (\ref{CM0}), note that this is defined using the probability distribution $P$ instead of $P^0$.

Finally, consider the distance between the average states sent to Bob in the ideal and practical protocols,
\begin{align}\label{Error_a}
\epsilon_\mathrm{a} := D(\sigma_A, \sigma_A^0) \, ,
\end{align}
where
\begin{align}
\sigma_A & = 
\sum_{j,k=0}^{d-1} \int_{J_j \times J_k} d^2\alpha P(\alpha) |\alpha_{jk} \rangle \langle \alpha_{jk} | \, , \\
\sigma_A^0 & =
\int d^2\alpha P^0(\alpha) |\alpha \rangle \langle \alpha | \, .
\end{align}

Note that in general $\epsilon_\mathrm{a}$, defined in Eq.\ (\ref{Error_a}), is smaller than $\epsilon_\mathrm{p}$, defined in Eq.\ (\ref{Error_p}). 
In Section \ref{Sec:example} we will see that in some cases $\epsilon_\mathrm{a}$ can be several orders of magnitude smaller than $\epsilon_\mathrm{p}$. 

It follows from the monotonicity property of the trace distance that $\epsilon_\mathrm{a}$ bounds the distance between the joint state of Bob and Eve, even after Bob's measurement and the application of the ADC, i.e.,
\begin{align}
D(\sigma_{YE}, \sigma_{YE}^0) & \leq \epsilon_\mathrm{a} \, , \\
D(\sigma_{\bar Y E}, \sigma_{\bar Y E}^0) & \leq \epsilon_\mathrm{a} \, , 
\label{barYE}
\end{align} 
as well as  
\begin{align} 
D(\sigma_Y, \sigma_Y^0)
= \frac{1}{2} \int d^2\beta | P(\beta) - P^0(\beta) | \leq \epsilon_\mathrm{a} \, ,
\label{distanceY}
\end{align} 
and $D(\sigma_{\bar Y}, \sigma_{\bar Y}^0) \leq \epsilon_\mathrm{a}$.
We will use the latter bounds to estimate the key rate in Section \ref{Sec:issue1} and the diagonal term in the covariance matrix in Section \ref{Sec:issue2}.

\begin{center}
\begin{table}
\begin{tabular}{ |c|l| } 
 \hline
Symbol & Meaning \\ 
\hline\hline
$R_A$ & Range of Alice's state preparation \\ 
$\delta_A$ & Bin size for Alice's state preparation \\ 
$M$ & Range of Bob's heterodyne measurement \\ 
$R_B$ & Range of Bob's ADC \\ 
$\delta_B$ & Bin size for Bob's ADC \\
$\log{d}, b$ & Bits per quadrature for Alice and Bob's raw key \\
$X$ & Alice's continuous variable \\
$\bar X$ & Alice's discrete variable \\
$Y$ & Bob's continuous variable \\
$\bar Y$ & Bob's discrete variable \\
$N$ & Mean photon number in signals sent from Alice \\
$n$ & Number of elementary transmissions (i.e.\ block size) \\
\hline
\end{tabular}
\caption{List of symbols used to characterize the protocol.}\label{Table:protocol}
\end{table}
\end{center}

\section{Main theoretical tools applied in this work}\label{Sec:toolbox}

Here I review the main theoretical tools applied for the security analysis presented in Section \ref{Sec:proof}.
The reader who is already familiar with these tools can skip this Section. 

{\it The leftover hash lemma} establishes a link between the min-entropy
of a random variable $\bar Y$ and the amount of uniform randomness that can be
extracted from it \cite{LHL}. 
Consider the bipartite state,
\begin{align}\label{exstate}
\rho_{\bar Y E} = \sum_y p_{\bar Y}(y) 
|y\rangle \langle y | \otimes \rho_E(y) \, ,
\end{align}
that describes the correlations between a classical random variable $\bar Y$ and a quantum system $E$. 
The latter may represent the quantum system under the control of the eavesdropper Eve.
Here $p_{\bar Y}(y)$ is the probability that $\bar Y$ takes value equal to $y$, and $\{ |y\rangle \}$ is a collection of orthogonal unit vectors that carry the values of $\bar Y$.
The min-entropy of $\bar Y$ conditioned on $E$, denoted as $H_\mathrm{min}(\bar Y|E)$, 
quantifies the probability of guessing the value of $\bar Y$ from measuring the system $E$.
In fact, the optimal probability of guessing is $p_\mathrm{guess} = 2^{-H_\mathrm{min}(\bar Y|E)}$.

According to the direct part of the leftover hash lemma, it is possible to extract from $\bar Y$ a string of $\ell^{\epsilon_\mathrm{h} + \epsilon_\mathrm{s}}$ random bits that are uniform and secret to Eve, up to a failure probability smaller than $\epsilon_\mathrm{h}+ \epsilon_\mathrm{s}$, where \cite{TR11,Tomamichel}
\begin{align}\label{lhl}
\ell^{\epsilon_\mathrm{h} + \epsilon_\mathrm{s}} \ge H_\mathrm{min}^{\epsilon_\mathrm{s}}(\bar Y|E) 
- 2 \log{(1/\epsilon_\mathrm{h})} 
+ 1 \, .
\end{align}
This bound is expressed in terms of the smooth min-entropy, $H_\mathrm{min}^{\epsilon_\mathrm{s}}(\bar Y|E)$, which is computed on a state $\rho_{\bar Y E}^{\epsilon_\mathrm{s}}$ that is $\epsilon_\mathrm{s}$-close to $\rho_{\bar Y E}$ \cite{Renner,Tomamichel}. 

{\it The Asymptotic Equipartition Property} (AEP) allows us to estimate
the smooth min-entropy in terms of the Shannon entropy. In fact, these two quantities coincides in the thermodynamic limit. 
For a $n$-fold tensor power of the state in Eq.\ (\ref{exstate}), the following bound holds \cite{Tomamichel}
\begin{align}\label{AEP0}
H_\mathrm{min}^{\epsilon_\mathrm{s}}(\bar Y^n|E^n) \geq n H(\bar Y|E) - \sqrt{n} \Delta(\epsilon_\mathrm{s},|\bar Y|) \, ,
\end{align}
where $\Delta(\epsilon_\mathrm{s},|\bar Y|)$ is a function of the smoothing parameter $\epsilon_\mathrm{s}$
and of the cardinality $|\bar Y|$ of the random variable $\bar Y$, 
with \cite{MDI_2018_1}
\begin{align}
\Delta(\epsilon_\mathrm{s},|\bar Y|) \leq 4\left( \frac{1}{2} \log{|\bar Y|} + 1 \right) \sqrt{ \log{\frac{2}{\epsilon_\mathrm{s}^2} }} \, .
\end{align}
Note that this expression for the finite-size correction in the AEP, obtained in Ref.\ \cite{MDI_2018_1}, is an improved version of the one of Ref.\ \cite{Lev1}. See Ref.\ \cite{MDI_2018_1} for more details.

To be precise, the state relevant to QKD applications is not necessarily a $n$-fold tensor power. This is due to the fact that a secret key is extracted only when Alice and Bob succeed in performing the error correction routine. Since error correction succeeds with non-unit probability, their joint state, conditioned on successful error correction, is no longer a tensor power \cite{Lev1}. Nevertheless, there are ways to circumvent this issue and still apply the AEP. A first approach to account for this issue was presented in Ref.\ \cite{Lev1}. A second approach was discussed in Ref.\ \cite{MDI_2018_1}.
According to Ref.\ \cite{MDI_2018_1}, we can account for the finite success probability of error correction simply by replacing $\Delta(\epsilon_\mathrm{s},|\bar Y|)$ with $\Delta\left( \tau p \epsilon_\mathrm{s},|\bar Y|\right)$, where $p$ is the probability of successful error correction and $\tau$ is a (small) constant. A feasible value is $\tau =\frac{2}{3}$ \cite{MDI_2018_1}.
Furthermore, an additive correction $\log{\left( p - \tau p \epsilon_\mathrm{s} \right)}$ has to be introduced.
This corrections are not considered here to keep the formulas as simple as possible. Including these corrections does not substantially change the results presented here.

{\it Continuity of the quantum mutual information.}
Several inequalities exist that bound the value of 
quantum entropies in the neighbour of a given quantum states,
see e.g.\ Refs.\ \cite{AF,KA,Winter2016,Shirokov}.
Here we apply an inequality for the quantum mutual information
as presented in Ref.\ \cite{Shirokov}.

Consider a pair of bipartite states, $\rho_{\bar Y E}$, $\rho_{\bar Y E}^0$, on quantum systems $\bar Y$ and $E$.
If $D(\rho_{\bar YE},\rho_{\bar YE}^0) \leq \epsilon$, then the following inequality holds:
\begin{align}\label{Continuity0}
\left| I(\bar Y;E)_\rho - I(\bar Y;E)_{\rho^0} \right| \leq f(\epsilon,|\bar Y|) \, ,
\end{align}
where $I(\bar Y;E)$ denotes the quantum mutual information, and \cite{Shirokov}
\begin{align}\label{ferror}
f(\epsilon,|\bar Y|) := 
\epsilon \log{|\bar Y|} + 2(1+\epsilon) \log{(1+\epsilon)} - 2\epsilon \log{\epsilon} \, .
\end{align}
Note that this bound is independent of the dimension of $E$,
which may be infinite.

{\it The optimality of Gaussian attacks} is a property of protocols where Alice prepares coherent states with a Gaussian amplitude distribution, and Bob measures by ideal homodyne or heterodyne detection. 
It establishes that, for given covariance matrix of Alice and Bob's quadratures, the optimal attack for Eve is a Gaussian attack. 
This is summarized by the inequality
\begin{align}\label{Gaussian0}
I(Y;E)_{\sigma^0} \leq I(Y;E)_{\sigma_G} \, ,
\end{align}
where $\sigma_G$ is a Gaussian state having the same
covariance matrix as $\sigma^0$,
and $Y$ denotes the random variable associated with the
outcome of ideal heterodyne measurement.

The important point that I want to emphasize here is that 
the theorem holds for the ideal protocol
where Alice prepares a Gaussian distribution of coherent
states and Bob applies ideal heterodyne. 
See Refs.\ \cite{Wolf,GarciaPatron,Navascues} for more detail.  

\section{A practical approach to assess the security of CV QKD}\label{Sec:proof}

In this Section, I outline an approach to assess the 
security of the protocol described in Section \ref{Sec:ideal},
assuming the particular practical implementation of Section \ref{Sec:practical}. The following subsections focus on imperfect state preparation and non-ideal heterodyne
detection.

The security analysis is developed within the framework of composable security \cite{Composable}. Therefore, the protocol is shown to be secure up to a small probability of error $\epsilon = \epsilon_\mathrm{h}+\epsilon_\mathrm{s}+\epsilon_\mathrm{a}$. The contributions to this and other errors parameters are summarized in Table \ref{Table:security}.

\begin{center}
\begin{table}
\begin{tabular}{ |c|l| } 
 \hline
Symbol & Meaning \\ 
\hline\hline
$\epsilon_\mathrm{h}$ & Hashing error \\
$\epsilon_\mathrm{s}$ & Entropy smoothing parameter \\
$\epsilon_\mathrm{p}$ & Symbol by symbol preparation error, Eq.\ (\ref{Error_p}) \\
$\epsilon_\mathrm{a}$ & Average preparation error, Eq.\ (\ref{Error_a}) \\
$\epsilon_{R_A,N}$ & $\sqrt{\frac{N}{2\pi}} \, e^{- \frac{R_A^2}{2N} }$ \\
\hline
\end{tabular}
\caption{List of symbols used in the security analysis.}\label{Table:security}
\end{table}
\end{center}

\subsection{The issue with discrete input modulation}\label{Sec:issue1}


For $n$ signal transmissions we are interested in finding a lower bound 
on the conditional smooth min-entropy
\begin{align}
H_\mathrm{min}^{\epsilon_\mathrm{s}}(\bar Y^n|E^n)_{\sigma^{\otimes n}} \, ,
\end{align}
where $\bar Y^n$ denotes $n$ instances of the Bob's discretized 
heterodyne output.
The min-entropy can then be used to bound the secret key rate
through the leftover hash lemma, see Section \ref{Sec:toolbox}.
As we assume collective attacks, the joint state of Bob and Eve is 
a tensor product, $\sigma^{\otimes n}$.

By applying the AEP we obtain, see Eq.\ (\ref{AEP0}),
\begin{align}
\frac{1}{n} H_\mathrm{min}^{\epsilon_\mathrm{s}}(\bar Y^n|E^n)_{\sigma^{\otimes n}} 
& \geq H(\bar Y|E)_\sigma - \frac{1}{\sqrt{n}} \, \Delta(\epsilon_\mathrm{s},|\bar Y|) \\
& \hspace{-1.5cm} = H(\bar Y)_\sigma - I(\bar Y;E)_\sigma
- \frac{1}{\sqrt{n}} \, \Delta(\epsilon_\mathrm{s},|\bar Y|) \, ,
\label{AEP1}
\end{align}
where the second equality follows from the identity $H(\bar Y|E)_\sigma = H(\bar Y)_\sigma - I(\bar Y;E)_\sigma$.
Note that, from the measurement data, Bob can empirically estimate the 
entropy $H(\bar Y)$, see Ref.\ \cite{Lev1}.




We now relate the mutual information $I(\bar Y;E)_\sigma$, which refers to the practical protocol, to the mutual information $I(\bar Y;E)_{\sigma^0}$, that would be obtained with the ideal protocol.
Using the condition in Eq.\ (\ref{barYE}) and the continuity bound on the mutual information in Eq.\ (\ref{Continuity0}), we obtain
\begin{align}
I(\bar Y;E)_\sigma \leq I(\bar Y ;E)_{\sigma^0} + f(\epsilon_\mathrm{a},|\bar Y|) \, .
\end{align}
Putting this in Eq.\ (\ref{AEP1}) we finally obtain
\begin{align}
\begin{split}
\frac{1}{n} H_\mathrm{min}^{\epsilon_\mathrm{s}}(\bar Y^n|E^n)_{\sigma^{\otimes n}} 
\geq H(\bar Y)_\sigma - I(\bar Y;E)_{\sigma^0} \\
- f(\epsilon_\mathrm{a},|\bar Y|)
- \frac{1}{\sqrt{n}} \, \Delta(\epsilon_\mathrm{s},|\bar Y|) 
 \, .
\end{split}
\label{AEP2}
\end{align}

A similar bound has been obtained Kaur {\it et al.}\ \cite{Wilde}. However, the bound of Ref.\ \cite{Wilde} depends on a free parameter that cannot be estimated experimentally.

%
%

\subsection{The issue with non-ideal heterodyne detection
and the optimality of Gaussian attacks}\label{Sec:issue2}

The next step is to relate the discrete and bounded variable $\bar Y$ 
with the continuous and unbounded variable $Y$ that would be obtained with
ideal heterodyne detection.
Note that the ADC transformation, $Y \to \bar Y$, defines 
a completely positive and trace-preserving map, therefore we can
apply the monotonicity property of the quantum mutual information to obtain
\begin{align}
I( \bar Y ; E)_{\sigma^0} \leq I( Y ; E)_{\sigma^0} \, .
\end{align}
Putting this in Eq.\ (\ref{AEP2}) we obtain
\begin{align}
\begin{split}
\frac{1}{n} H_\mathrm{min}^{\epsilon_\mathrm{s}}(\bar Y^n|E^n)_{\sigma^{\otimes n}} 
\geq H(\bar Y)_\sigma - I(Y;E)_{\sigma^0} \\
- f(\epsilon_\mathrm{a},|\bar Y|)
- \frac{1}{\sqrt{n}} \, \Delta(\epsilon_\mathrm{s},|\bar Y|) 
 \, .
\end{split}
\label{AEP3}
\end{align}

We can now apply the property of optimality of Gaussian attacks.
In fact, the mutual information $I(Y;E)_{\sigma^0}$ is defined for the
variable $Y$ that is the output of ideal heterodyne detection, and is computed
on the state $\sigma^0$ that is generated for the ideal state preparation.
We can then insert Eq.\ (\ref{Gaussian0}) into (\ref{AEP3}) and obtain
\begin{align}\label{ratefinal}
\begin{split}
\frac{1}{n} H_\mathrm{min}^{\epsilon_\mathrm{s}}(\bar Y^n|E^n)_{\sigma^{\otimes n}} 
\geq H(\bar Y)_\sigma - I(Y;E)_{\sigma_G} \\
- f(\epsilon_\mathrm{a},|\bar Y|)
- \frac{1}{\sqrt{n}} \, \Delta(\epsilon_\mathrm{s},|\bar Y|)  \, ,
\end{split}
\end{align}
where $\sigma_G$ is a Gaussian state that has the same covariance 
matrix as $\sigma^0$.

Unlike $I(Y;E)_{\sigma^0}$, the Gaussian mutual information
$I(Y;E)_{\sigma_G}$ is uniquely determined by the
covariance matrix of $\sigma^0$, i.e., the quantity defined in
Eq.\ (\ref{CM0}).
The problem is that the state $\sigma^0$ is neither prepared nor measured in the laboratory. The only state that is physically accessible is $\sigma$.
So we are in the position of having to estimate the covariance matrix of
$\sigma^0$ from measuring $\sigma$.

The fact that these two states are close in trace distance is, in general, not sufficient to bound the difference between their covariance matrices, see Ref.\ \cite{Wilde}. This is essentially due to the fact that the variable $Y$ is unbounded.
Kaur \textit{et al.}\ addressed this problem by introducing two parameters, called $c_1$ and $c_2$, that quantify the distance between the state $\sigma$ and $\sigma^0$. Although the final result only mildly depends on these parameters, there is no known way to determine $c_1$ and $c_2$.

Here I do not solve this problem, but propose a practical workaround that takes into account the data that is actually collected in the laboratory. My approach is not mathematically rigorous, but it is physically motivated.
%
It is based on the observation that the range of any experimental realization of heterodyne is limited by saturation and non-linear effects.
This means that Bob can only measure values of the quadratures $q_B$, $p_B$ within a finite range $\mathcal{M} = [-M,M] \times [-M,M]$.
As a matter of fact, this is what is necessarily done in any experimental realization of CV QKD.

From the physical point of view, Eqs.\ (\ref{AEP3})-(\ref{ratefinal}) are not very informative, as they are written in terms of the covariance matrix of ideal heterodyne, which cannot be estimated experimentally. Therefore, I will address the physically well-defined question: {\it What would be the covariance matrix of non-ideal heterodyne on the state $\sigma^0$?}
It is then possible to compare the covariance matrix of non-ideal heterodyne detection on the states $\sigma$ and $\sigma^0$.

Consider for example the quadrature $q_B$ and compare
the mean values $\langle q_B^2 \rangle_{P}$ and 
$\langle q_B^2 \rangle_{P^0}$, which refer to $\sigma$ and $\sigma^0$, respectively.
We have
\begin{align}
\left| \langle q_B^2 \rangle_{P} - \langle q_B^2 \rangle_{P^0} \right|
%
%
& \leq \int_\mathcal{M} \left| P(\beta) - P^0(\beta) \right| q_B^2 d^2\beta \\ 
& \leq M^2 \int_\mathcal{M} \left| P(\beta) - P^0(\beta) \right| d^2\beta \\
& \leq 2 \epsilon_\mathrm{a} M^2 \, ,
\label{qb2}
\end{align}
where the second inequality follows from the fact that $q_B \in [-M,M]$, and the last inequality follows form Eq.\ (\ref{distanceY}).
The same bound can be obtained for the quadrature $p_B$.

Consider now the cross-diagonal terms, for example $\langle q_A q_B \rangle_P$ and $\langle q_A q_B \rangle_{P^0}$. 
We have:
\begin{align}
& \left| \langle q_A q_B \rangle_{P} - \langle q_A q_B \rangle_{P^0} \right| \nonumber \\
%
%
& \hspace{1cm} \leq \int d^2\alpha \int_\mathcal{M} d^2\beta  \left| P(\alpha,\beta) - P^0(\alpha,\beta) \right| \left| q_A q_B \right| \\
& \hspace{1cm} \leq M \int d^2\alpha \int_\mathcal{M} d^2\beta \left| P(\alpha,\beta) - P^0(\alpha,\beta) \right| \left| q_A \right| \\
& \hspace{1cm} \leq R_A M \int_{q_A\leq R_A} \hspace{-0.5cm} d^2\alpha \int d^2\beta \left| P(\alpha,\beta) - P^0(\alpha,\beta) \right| \nonumber \\
& \hspace{1cm} + M \int_{q_A>R_A} \hspace{-0.5cm} d^2\alpha \int d^2\beta P^0(\alpha,\beta) \left| q_A \right| \label{Mlarger} \\
& \hspace{1cm} \leq 2 R_A M \epsilon_\mathrm{p}
+ M \, \epsilon_{R_A,N} \, ,
\label{last}
\end{align}
where inequality (\ref{Mlarger}) follows from the fact that
$P(\alpha,\beta) = 0$ if $|q_A|>R_A$,
and the last inequality follows from Eq.\ (\ref{Pdistance}). 
Furthermore, in the last inequality we have introduced the notation
\begin{align}
\epsilon_{R_A,N} & := \int_{q_A>R_A} \hspace{-0.5cm} d^2\alpha \int d^2\beta P^0(\alpha,\beta) \left| q_A \right| \\
& =
\int_{q_A>R_A} d^2\alpha P^0(\alpha) \left| q_A \right| \\
& = \sqrt{\frac{N}{2\pi}} \, e^{- \frac{R_A^2}{2N} } \, ,
\end{align}
which builds on Eq.\ (\ref{Gaussian}).
In a similar way we can bound all the other cross-diagonal terms,
$\langle q_A p_B \rangle_{P^0}$,
$\langle p_A q_B \rangle_{P^0}$, and
$\langle p_A p_B \rangle_{P^0}$.

Note that the signal-by-signal preparation error $\epsilon_\mathrm{p}$ appears in the estimate of the cross-diagonal terms, whereas the correction in the diagonal terms only depends on the average preparation error $\epsilon_\mathrm{a}$.

\section{Extension to coherent attacks}\label{Sec:coherent}

To establish the security against most general coherent attacks is the final goal of the security analysis of any QKD protocols. Due to the complexity of the problem, the security proofs are typically built step after step. First focusing on collective attacks, and then seeking an extension to coherent attacks. For the case of CV QKD protocols with discrete input modulation, asymptotic security proofs were obtained in 2019 in Refs.\ \cite{LevPRX,NL2}, and still it is not known how these results can be extended to include finite-size effects and coherent attacks.

Here I have presented an approach that is independent and complementary to those of Refs.\ \cite{LevPRX,NL2}, focusing on collective attacks in the non-asymptotic regime. An extension to coherent attacks is still premature at this stage because, as discussed above, this approach is not yet able to provide a full and rigorous security proof for collective attacks.

The post-selection technique has been applied successfully to extend the security of QKD protocols from collective to coherent attacks \cite{post-s}. This technique exploits the symmetries of the protocol. Typically, CV QKD protocols are symmetric under permutation of the signal transmissions \cite{Renner_Cirac_2009}, and under multi-mode unitary transformations that mix many signal transmissions \cite{Lev2013,Lev2}. Discrete modulation breaks the symmetry under unitary transformations, therefore, if the post-selection technique is to be applied, one should consider other symmetries.
Another ingredient of known security proof against coherent attacks is an energy test that is used the effective dimension of the system. In principle, the energy test does not depend on the symmetry of the protocol, and therefore can be applied to discrete-modulation CV QKD protocols.

\section{Examples}\label{Sec:example}

Consider an example where Alice samples the coherent state amplitudes from a regular square lattice of size $2^n$, i.e., encoding $n$ bits per quadrature.
This encoding covers a region in phase space of length $2R_A$.
Following the general procedure described in Section \ref{Sec:practical}, this region is divided in intervals $J_j \times J_k$, with
\begin{align}
J_j = ( - R_A + j\delta_A , - R_A + (j+1)\delta_A ] \, , 
\end{align}
for $j=0,\dots, 2^n-1$, 
with $\delta_A = 2R_A/2^n$.
To each interval is associated a unique complex number
$\alpha_{jk} = ( q_{Aj} + ip_{Ak} )/\sqrt{2}$, 
where $q_{Aj} = - R_A + j\delta_A + \delta_A/2$ 
and $p_{Ak} = - R_A + k\delta_A + \delta_A/2$.

\begin{figure}[t]
\centering
\includegraphics[width=0.5\textwidth]{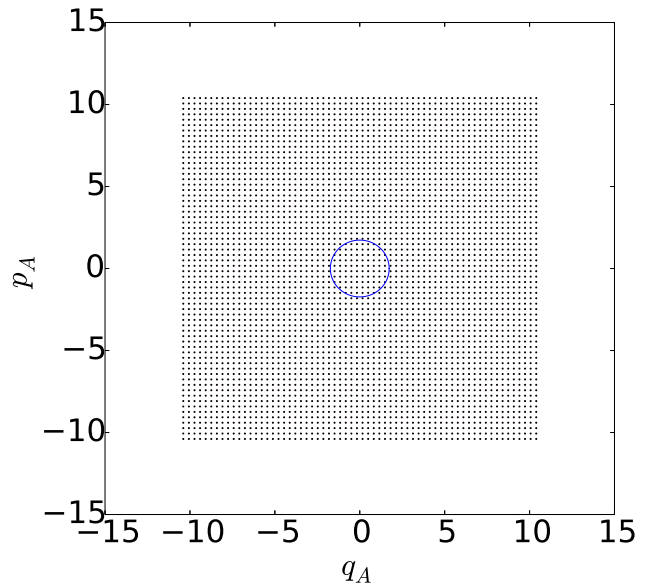}
\caption{A grid of coherent state amplitudes in phase space. The parameters are: $b = 6$ (i.e., $6$ bits per quadrature), $R_A = 6 \sqrt{N}$ (i.e., the range covers $6$ standard deviations), and the mean photon number is $N = 3$.
The blue circle has radius equal to the standard deviation $\sqrt{N}$.
The average preparation error is $\epsilon_\mathrm{a} \leq 10^{-6}$.
The value $N=3$ is similar to the value used, for example, in Ref.\ \cite{MLCVQKD}), and the values of $b=6$ bits per quadrature and of $6$ standard deviations are similar to the values used in Ref.\ \cite{Jouguet}.}
\label{Fig:grid}
\end{figure}

To provide a visual intuition to the reader, Fig.\ \ref{Fig:grid} shows the grid of coherent states generated in this way, for $b = 6$ (i.e., $6$ bits per quadrature), and $R_A = 6 \sqrt{N}$ (i.e., the range covers $6$ standard deviations), and the mean photon number is $N = 3$. (These values for the parameters are in the same range as the parameters used in Refs.\  \cite{Jouguet,MLCVQKD}.)

The amplitudes are sampled from the probability distribution \cite{Jouguet}
\begin{align}
P(\alpha) = \mathcal{N}
\chi_{J_j \times J_k} e^{-|\alpha_{jk}|^2/N} \, ,
\end{align}
where $\chi_{J_j \times J_k}$ is the characteristic function of $J_j \times J_k$, and $\mathcal{N}$ is a normalisation factor.
The average preparation error can be estimated numerically. Following the recipe of Ref.\ \cite{Jouguet}, and using the values of $b$, $R_A$, and $N$ given above, we obtain $\epsilon_\mathrm{a} \leq 10^{-6}$.

\begin{figure}[t]
\centering
\includegraphics[width=0.5\textwidth]{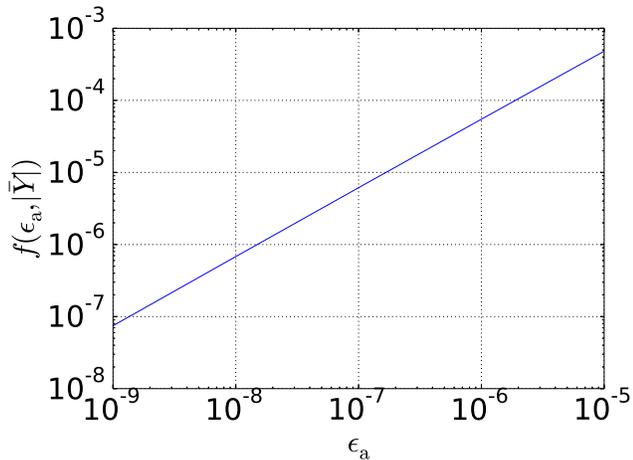}
\caption{This shows the additive error $f(\epsilon_\mathrm{a},|\bar{Y}|)$ in the key rate (see Eq.\ (\ref{keyrateE1})) versus the average preparation error $\epsilon_\mathrm{a}$ (see Eq.\ (\ref{Error_a})). 
The plot is obtained for $|\bar{Y}| = 2^{12}$, i.e., $6$ bits per quadrature. Note that $f(\epsilon_\mathrm{a},|\bar{Y}|)$ grows only logarithmically with $|\bar{Y}|$.}
\label{Fig:f}
\end{figure}

The error parameter $\epsilon_\mathrm{a}$ is relevant to estimate the secret key rate through the quantum mutual information. 
Putting Eq.\ (\ref{ratefinal}) into (\ref{lhl}) we obtain the following estimate of the secret key rate:
\begin{align}\label{keyrateE1}
\begin{split}
r^{\epsilon_\mathrm{h} + \epsilon_\mathrm{s} }_n =
H(\bar Y)_\sigma - I(Y;E)_{\sigma_G} - \mathrm{leak_{EC}} \\
- f(\epsilon_\mathrm{a},|\bar Y|) 
- \frac{1}{\sqrt{n}} \, \Delta(\epsilon_\mathrm{s},|\bar Y|) \, ,
\end{split}
\end{align}
where $\mathrm{leak_{EC}}$ is the number of bits leaked for error reconciliation, and we have neglected the terms 
$- \frac{2}{n} \log{(1/\epsilon_\mathrm{h})}
+ \frac{1}{n}$, as this is of higher order in $n$.
The correction term $f(\epsilon_\mathrm{a},|\bar Y|)$ is plotted in Fig.\ \ref{Fig:f}, showing that it grows linearly with $\epsilon_\mathrm{a}$.

For the sake of comparison, we put 
\begin{align}
H(\bar Y)_\sigma - \mathrm{leak_{EC}} \equiv \beta I(X;Y)_{\sigma_G} \, ,
\end{align}
where $\beta < 1$ is the error correction efficiency, and
$I(X;Y)_{\sigma_G}$ is the mutual information between Alice and Bob for an ideal protocol
with Gaussian modulation and under Gaussian attack.
We further assume an entangling cloner attack with loss factor $\eta$ and excess noise $u = (1-\eta) \omega$.
This is equivalent to say that the channel from Alice to Bob is a thermal loss channel.
Table \ref{Table:channel} summarizes the parameters that characterize this channel model.
This yields the asymptotic key rate
\begin{align}
r_\infty & = \beta I(X;Y)_{\sigma_G} - I(Y;E)_{\sigma_G} \\
& = \beta \log{\left[ 1 + \frac{\eta N}{(1-\eta)\omega + 1} \right]}
- g[N] + g[(1-\eta) \tilde N] \, ,
\label{rinfty}
\end{align}
where $g[x] := (x+1)\log{(x+1)} - x \log{x}$, and 
\begin{align}
\tilde N = \frac{N(1+\omega)}{ 1 + \eta N + (1-\eta) \omega } \, .
\end{align}

\begin{figure}[t]
\centering
\includegraphics[width=0.5\textwidth]{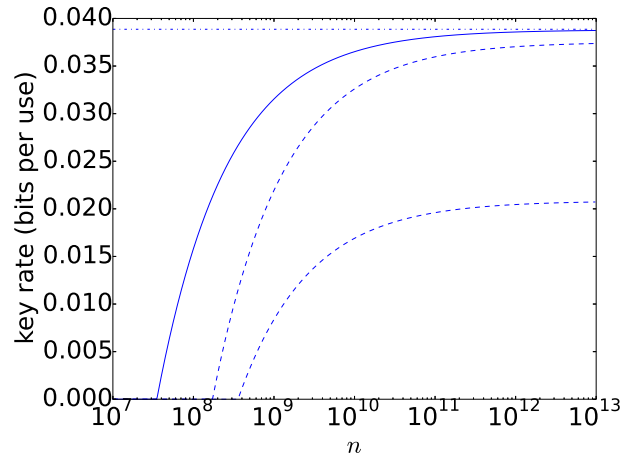}
\caption{The plot shows the key rate in bits per channel use, for loss $\eta = 0.1$ and excess noise $u = 10^{-4}$.
The dot-dashed line shows the ideal, asymptotic rate $r_\infty$ in Eq.\ (\ref{rinfty}). 
The solid line shows the non-ideal, non-asymptotic estimate for the secret key in Eq.\ (\ref{keyrate01}), with $\epsilon_\mathrm{s}=10^{-10}$,
$\epsilon_\mathrm{a}=10^{-6}$, and
$|\bar Y| = 2^{12}$ ($b=6$ bits per quadrature).
The symbol-by-symbol error parameter $\epsilon_\mathrm{p}$ is virtually put equal to $0$.
The main correction term in the solid line is due to the term $\frac{1}{\sqrt{n}} \, \Delta(\epsilon_\mathrm{s},|\bar Y|)$.
The dashed lines are obtained by also including the additive error in the estimation of the covariance matrix. From bottom to top, with $b=11$ and $b=14$ bits per quadrature.
Note that more bits per quadrature are needed to have a non-zero key rate because the additive error is dominated by the symbol-by-symbol error $\epsilon_\mathrm{p}$.}
\label{Fig:key}
\end{figure}

Figure \ref{Fig:key} (solid line) shows the key rate versus the block size $n$,
\begin{align}
r^{\epsilon_\mathrm{h} + \epsilon_\mathrm{s}}_n =
r_\infty
- f(\epsilon_\mathrm{a},|\bar Y|) 
- \frac{1}{\sqrt{n}} \, \Delta(\epsilon_\mathrm{s},|\bar Y|) \, ,
\label{keyrate01}
\end{align}
and compares it with the asymptotic rate $r_\infty$ (dot-dashed line). 
The figure is obtained for 
$\epsilon_\mathrm{s}=10^{-10}$,
$\epsilon_\mathrm{a}=10^{-6}$, and
$|\bar Y| = 2^{12}$ ($6$ bits per quadrature).
With this choice of parameters we have 
$f(\epsilon_\mathrm{a},|\bar Y|) \ll
\frac{1}{\sqrt{n}} \, \Delta(\epsilon_\mathrm{s},|\bar Y|)$.

Let us now look at the estimation of the covariance matrix. 
Figure \ref{Fig:key} has been obtained assuming given values of the channel parameters: loss and excess noise. 
These parameters are in turn estimated from the covariance matrix of the quadratures.
However, the discussion in Section \ref{Sec:issue2} showed that there could be a difference between the covariance matrix of the ideal and real protocol, the difference between the entries of the covariance matrix being bounded as in Eq.\ (\ref{qb2}) for the diagonal terms, and Eq.\ (\ref{last}) for the off-diagonal ones.

The error in the diagonal terms is determined by the parameter $\epsilon_\mathrm{a}$.
With a preparation routine based on $6$ bits per quadrature, $6$ standard deviation, and mean photon number $N=3$, we obtain $\epsilon_\mathrm{a} \leq 10^{-6}$.
Assuming a range for Bob's heterodyne of $6$ standard deviations, i.e., 
\begin{align}
M \simeq 6 \sqrt{\eta N + u} \, , 
\end{align}
the error in the diagonal terms is 
\begin{align}
2 \epsilon_\mathrm{a} M^2 \simeq 72 \times 10^{-6} (\eta N + u) < 10^{-4} (\eta N + u) \, .
\end{align}
Given that the expected value of the diagonal terms of the covariance matrix is $\eta N + u$, this yields a relative error smaller than $10^{-4}$.

The error in the off-diagonal terms of the covariance matrix is instead determined by $\epsilon_\mathrm{p}$. 
This is problematic because the latter can be much larger than $\epsilon_\mathrm{a}$.
In fact, combining Eq.\ (\ref{idealXA}) and Eq.\ (\ref{realXA}) into Eq.\ (\ref{Error_p}), we obtain
\begin{align}
\epsilon_{p} & =
\frac{1}{2} \sum_{j,k=0}^{d-1} \int_{J_j \times J_k} d^2\alpha \| P(\alpha) |\alpha_{jk} \rangle \langle \alpha_{jk} | 
- P^0(\alpha) |\alpha \rangle \langle \alpha| \| \\
& \simeq 
\frac{1}{2} \sum_{j,k=0}^{d-1} \int_{J_j \times J_k} d^2\alpha P^0(\alpha) \| |\alpha_{jk} \rangle \langle \alpha_{jk} | 
-  |\alpha \rangle \langle \alpha| \| \\
& = 
\sum_{j,k=0}^{d-1} \int_{J_j \times J_k} d^2\alpha P^0(\alpha) \sqrt{ 1 - e^{-|\alpha-\alpha_{jk}|^2}} \\
& \simeq
\sum_{j,k=0}^{d-1} \int_{J_j \times J_k} d^2\alpha P^0(\alpha) |\alpha-\alpha_{jk}| \, .
\end{align}
The first approximation comes from the fact that $P(\alpha) \simeq P^0(\alpha)$, the second equality follows from the expression of the trace distance between coherent states, and the last approximation is a first order Taylor expansion in $|\alpha-\alpha_{jk}|$.

Note that
\begin{align}
|\alpha-\alpha_{jk}| & =
\frac{1}{\sqrt{2}} \sqrt{ ( q_A - {q_A}_j )^2 + ( p_A - {p_A}_j )^2 } \, .
\end{align}
Therefore, given a preparation routine with range $R_A$ and $b$ bits per quadrature, then $\delta_A = \frac{2R_A}{2^b}$ is the bin size, which yields 
\begin{align}
q_A - {q_A}_j & \simeq \frac{\delta_A}{2} \, , \\
p_A - {p_A}_j & \simeq \frac{\delta_A}{2} \, ,
\end{align}
and thus
\begin{align}
|\alpha-\alpha_{jk}| \simeq  \frac{\delta_A}{2}
\, .
\end{align}
In conclusions we obtain the following estimate for $\epsilon_{p}$,
\begin{align}
\epsilon_{p} & \simeq \frac{\delta_A}{2}
\, .
\end{align}

This shows that the signal-by-signal preparation error $\epsilon_{p}$ is determined by Alice's bin size only.
For example, using $b = 6$ bits per quadrature, $6$ standard deviation, and mean photon number $N=3$, we obtain 
\begin{align}
\delta_A = 2 \times 6 \sqrt{N}/2^6 \simeq 0.32 \, ,
\end{align}
and $\epsilon_{p} \simeq 0.16$, which is about $5$ orders of magnitude larger than $\epsilon_\mathrm{a}$!

A relatively large value of $\epsilon_\mathrm{p}$ has a negative impact on the estimation of the off-diagonal terms in the covariance matrix. 
From Eq.\ (\ref{last}), putting $R_A = 6 \sqrt{N}$ and $M = 6 \sqrt{ \eta N + u} \simeq 6 \sqrt{ \eta N}$, we obtain an error of the order of
\begin{align}
2 R_A M \epsilon_\mathrm{p} + M \epsilon_{R_A,N} 
& \simeq 72 \sqrt{\eta} N \epsilon_\mathrm{p} + 6 \sqrt{\eta} N \frac{e^{-18}}{\sqrt{2\pi}} \\
& \simeq \left( 72 \epsilon_\mathrm{p} + 10^{-8} \right) \sqrt{\eta} N  \, .
\end{align}
Therefore, to make this error term sufficiently small, we need $\epsilon_\mathrm{p}$ to be much smaller than $1/72\simeq 0.014$.
In turn, with $R_A = 6 \sqrt{N}$ and $N=3$, to obtain $\epsilon_\mathrm{p} = 1/72$ we need to put $b > 11$, i.e., use at least $11$ bits per quadrature.

Figure \ref{Fig:key} (dashed lines) shows the key rate obtained by taking into account the additive errors to the covariance matrix.
The two dashed lines in Fig.\ \ref{Fig:key} are obtained for $b=11$ and $b=14$, i.e., $11$ and $14$ bits per quadrature.
The figure shows how the key rate drops if $b$ is not large enough.

\begin{center}
\begin{table}
\begin{tabular}{ |c|l| } 
 \hline
Symbol & Meaning \\ 
\hline\hline
$\eta$ & Loss factor \\
$u$ & Excess noise, $u = (1-\eta) \omega$. \\
$\omega$ & Eve's mean photons in the entangled cloner attack. \\
\hline
\end{tabular}
\caption{List of symbols used to characterize the channel parameters.}\label{Table:channel}
\end{table}
\end{center}

\section{Summary of results}\label{Sec:summary}

Here I have considered two main discrepancies between the experimental implementations of CV QKD and the mathematical models used to prove its security. They account for imperfections in state preparation and measurements. 
In summary:

\begin{enumerate}

\item To deal with imperfect state preparation, I have exploited the continuity of the quantum mutual information. 
Given an imperfect state preparation routine with average preparation error $\epsilon_\mathrm{a}$, the key rate differs from that of the ideal protocol by an additive term given in Eq.\ (\ref{ferror}),
\begin{align}
\begin{split}
f(\epsilon_\mathrm{a},|\bar Y|) & =
\epsilon_\mathrm{a} \log{|\bar Y|} \\
& + 2(1+\epsilon_\mathrm{a}) \log{(1+\epsilon_\mathrm{a})} 
- 2\epsilon_\mathrm{a} \log{\epsilon_\mathrm{a}} \, ,
\end{split}
\end{align}
see Eq.\ (\ref{AEP2}), where, in reverse reconciliation, $|\bar Y|$ is the cardinality of Bob's raw key.

A similar bound, previously proposed in Ref.\ \cite{Wilde}, was expressed in term of a function $\tilde f(\epsilon_\mathrm{a},P_E)$. 
Note that the latter depends on the parameter $P_E$, which quantifies the mean photon number of the eavesdropper. 
Unfortunately, there is no known way for Alice and Bob to estimate $P_E$.



\item Using the continuity of the quantum mutual information requires to estimate the covariance matrix of a state that is not physically accessible.
I have in part resolved this ambiguity by invoking a physical argument based on the fact any experimental realization of heterodyne detection has necessarily a finite range.
This has allowed me to bound the covariance matrix of the unphysical state with that of a physically accessible one.
This approach should be considered as a practical workaround and has no ambitious of mathematical rigour.

For the diagonal terms of the covariance matrix, for example $\langle q_B^2 \rangle$, this bound has the form, see Eq.\ (\ref{qb2}),
\begin{align}
\langle q_B^2 \rangle_\mathrm{unphys} \leq 
\langle q_B^2 \rangle_\mathrm{phys} +
2 M^2 \epsilon_\mathrm{a} \, .
\end{align}
For the off-diagonal terms, for example $\langle q_A q_B \rangle$, we obtain, see Eq.\ (\ref{last}),
\begin{align}
& \langle q_A q_B \rangle_\mathrm{unphys} \geq \nonumber \\
& \hspace{0.6cm} 
\langle q_A q_B \rangle_\mathrm{phys} 
- 2 R_A M \epsilon_\mathrm{p}
- M \, \sqrt{\frac{N}{2\pi}} \, e^{- \frac{R_A^2}{2N} } \, .
\end{align}
For the meaning of the parameters $M$, $R_A$, $N$, please refer to Table \ref{Table:protocol}.

\item The analysis of the physical implementations of the CV QKD protocol has been developed in terms of two preparation errors, $\epsilon_\mathrm{a}$ and $\epsilon_\mathrm{p}$, both defined as a trace distances:
\begin{itemize}

\item $\epsilon_\mathrm{a}$, defined in Eq.\ (\ref{Error_a}), is the average error in the state preparation. This quantity was also considered in other works, most notably in Ref.\ \cite{Jouguet};

\item $\epsilon_\mathrm{p}$, defined in Eq.\ (\ref{Error_p}), is a symbol-by-symbol error. This can be orders of magnitude larger than $\epsilon_\mathrm{a}$. 

\end{itemize}

The analysis presented here shows that $\epsilon_\mathrm{p}$, and not $\epsilon_\mathrm{a}$, is the relevant parameter for the estimation of the off-diagonal elements of the covariance matrix.
To the best of my knowledge, this fact was not noticed in previous literature.
%

\end{enumerate}

\section{Conclusions}\label{Sec:conclusions}

This work has explored the gap between theoretical security proofs and experimental implementations of CV QKD.
Unlike other works, I have made use of both mathematical tools and physical arguments to assess the security of a practical CV QKD protocol.
The result is a security analysis that is as rigorous as it can be, given the theoretical tools we have in our toolbox and the limitations of experimental practice.
This approach indeed mirrors what is done in practice in laboratory realizations of CV QKD.
I have focused on the one-way protocol of Ref.\ \cite{no-switching}, but similar conclusions can be drawn for two-way \cite{Lev2way,2way} and Measurement-Device-Independent protocols \cite{MDI,MDI_2018_2}.

I have discussed some of the discrepancies between mathematical models and experimental realizations of CV QKD. 
On one hand, we have elegant mathematical theorems that exploit symmetries in infinite-dimensional Hilbert spaces, see e.g.\ Refs.\ \cite{Lev1,Lev2,GarciaPatron,GaussDeFinetti}.
On the other hand, physicists and engineers deal with imperfect and noisy devices and operate with limited resources.
%

In principle, one expects the mathematical models to be meaningful approximations of physical reality, whose scope is to guide the experimental work.
In CV QKD, I see the risk that the experimenter might be faced with the overwhelming task of having to reproduce the mathematical models with unrealistic levels of confidence. 
If this is the case, the models may lose their usefulness to physics.
This work represents an attempts to highlight and mitigate this risk.

In conclusions, this work has outlined an experimentally-friendly approach to assess the security of CV QKD. 
More effort is necessary to develop this into a full security proof (for example, I have not discussed here the routine of parameter estimation, which, however, has been studied extensively in other works, see, e.g.\ \cite{Lev1,Lev2010,Usenko}).
This contribution goes in the same direction of other recent works that have discussed protocols and security proofs that, by definition, require a finite constellation of input coherent states \cite{LevPRX,NL2,NL1,Kamil,PP,Wilde}. 


\section*{Acknowledgements}

This work was supported by the EPSRC Quantum Communications Hub, Grant  No.EP/M013472/1.
I acknowledge numerous insightful discussions with Tobias Gehring and Anthony Leverrier.


\begin{thebibliography}{99}

\bibitem{Scarani2009}
V. Scarani, H. Bechmann-Pasquinucci, N. J. Cerf, M. Dusek, N. Lutkenhaus, M. Peev,
Rev. Mod. Phys. {\bf 81}, 1301 (2009).

\bibitem{Diamanti}
E. Diamanti, H.-K. Lo, B. Qi and Z. Yuan, 
npj Quantum Information {\bf 2}, 16025 (2016).

\bibitem{Bacco}
D. Cozzolino, B. Da Lio, D. Bacco, L. K. Oxenløwe,
Advanced Quantum Technologies {\bf 2}, 1900038 (2019).

\bibitem{EReview}
E. Diamanti, A. Leverrier,
Entropy {\bf 17}, 6072 (2015).

\bibitem{Furrer}
F. Furrer, T. Franz, M. Berta, A. Leverrier, V. B. Scholz, M. Tomamichel, R. F. Werner,
Phys. Rev. Lett. {\bf 109}, 100502 (2012); 
Phys. Rev. Lett. {\bf 112}, 019902(E) (2014).

\bibitem{FurrerRR}
F. Furrer, 
Phys. Rev. A {\bf 90}, 042325 (2014). 

\bibitem{Lev1}
A. Leverrier, 
Phys. Rev. Lett. \textbf{114}, 070501 (2015).

\bibitem{Lev2}
A. Leverrier, 
Phys. Rev. Lett. \textbf{118}, 200501 (2017).

\bibitem{Lev2way}
S. Ghorai, E. Diamanti, and A. Leverrier,
Phys. Rev. A {\bf 99}, 012311 (2019).

\bibitem{no-switching}
C. Weedbrook, A. M. Lance, W. P. Bowen, T. Symul, T. C. Ralph, and P. K. Lam,
Phys. Rev. Lett. {\bf 93}, 170504 (2004).

\bibitem{Jouguet}
P. Jouguet, S. Kunz-Jacques, E. Diamanti, and A. Leverrier,
Phys. Rev. A {\bf 86}, 032309 (2012).

\bibitem{Wilde}  
E. Kaur, S. Guha, and M. M. Wilde,
arXiv:1901.10099 (2019).

\bibitem{Jouguet2}
P. Jouguet, D. Elkouss, S. Kunz-Jacques,
Phys. Rev. A {\bf 90}, 042329 (2014).

\bibitem{Jouguet3}
P. Jouguet, S. Kunz-Jacques, A. Leverrier, P. Grangier, E. Diamanti,
Nature Photonics {\bf 7}, 378 (2013).

\bibitem{Jouguet4}
P. Jouguet, S. Kunz-Jacques, T. Debuisschert, S. Fossier, E. Diamanti, R. All\'eaume, R. Tualle-Brouri, P. Grangier, A. Leverrier, P. Pache, P. Painchault,
Opt. Express {\bf 20}, 14030 (2012).

\bibitem{MDI_2018_1}
C. Lupo, C. Ottaviani, P. Papanastasiou, S. Pirandola,
Phys. Rev. A {\bf 97}, 052327 (2018).

\bibitem{GarciaPatron}
R. Garc\'ia-Patr\'on and N. J. Cerf, 
Phys. Rev. Lett. {\bf 97}, 190503 (2006).

\bibitem{Navascues} 
M. Navascu\'es, F. Grosshans, and A. Ac\'in, 
Phys. Rev. Lett. {\bf 97}, 190502 (2006).

\bibitem{LHL}
R. Impagliazzo, L. A. Levin, M. Luby,
Proceedings of the 21st Annual ACM Symposium on Theory of Computing, May 14-17, 1989, Seattle, Washington, USA, pp. 12-24.

\bibitem{TR11} M. Tomamichel, C. Schaffner, A. Smith, R. Renner.
IEEE Trans. on Inf. Theory {\bf 57}, 5524 (2011).

\bibitem{Tomamichel}
M. Tomamichel,
Ph.D. thesis,
Swiss Federal Institute of Technology (ETH) Zurich, 2012, 
arXiv:1203.2142 (2012).

\bibitem{Renner}
R. Renner, 
Ph.D. thesis, 
Swiss Federal Institute of Technology (ETH) Zurich, 2005, 
arXiv:0512258 (2005).

\bibitem{AF}
R. Alicki, M. Fannes, 
J. Phys. A: Math. Gen. {\bf 37}, L55 (2004).

\bibitem{KA}
K. M. R. Audenaert, 
J. Phys. A: Math. Gen. {\bf 40}, 8127 (2007).

\bibitem{Winter2016}
A. Winter,
Commun. Math. Phys. {\bf 347}, 291 (2016).

\bibitem{Shirokov}
M. E. Shirokov,
J. Math. Phys. {\bf 58}, 102202 (2017).

\bibitem{Wolf}  
M. M. Wolf, G. Giedke, and J. I. Cirac,
Phys. Rev. Lett. {\bf 96}, 080502 (2006).

\bibitem{Composable}
M. Ben-Or, M. Horodecki, D. W. Leung, D. Mayers, J. Oppenheim,
Theory of Cryptography: Second Theory of Cryptography Conference, TCC 2005, J.Kilian (ed.) Springer Verlag 2005, vol. 3378 of Lecture Notes in Computer Science, pp. 386-406.

\bibitem{LevPRX}
S. Ghorai, P. Grangier, E. Diamanti, and A. Leverrier,
Phys. Rev. X {\bf 9}, 021059 (2019).

\bibitem{NL2}
J. Lin, T. Upadhyaya, N. L\"utkenhaus,
Phys. Rev. X {\bf 9}, 041064 (2019).

\bibitem{post-s}
M. Christandl, R. Koenig, R. Renner,
Phys. Rev. Lett. \textbf{102}, 020504 (2009).

\bibitem{Renner_Cirac_2009}
R. Renner and J. I. Cirac, 
Phys. Rev. Lett. {\bf 102}, 110504 (2009).

\bibitem{Lev2013}
A. Leverrier, R. Garc\'{i}a-Patr\'{o}n, R. Renner, N. J. Cerf.
Phys. Rev. Lett. \textbf{110}, 030502 (2013).

\bibitem{MLCVQKD}
H.-M. Chin, N. Jain, D. Zibar, U. L. Andersen, T. Gehring
arXiv:2002.09321 (2020).

\bibitem{2way}
C. Ottaviani, S. Mancini, S. Pirandola,
Phys. Rev. A {\bf 92}, 062323 (2015).

\bibitem{MDI}
S. L. Braunstein, S. Pirandola,
Phys. Rev. Lett. {\bf 108}, 130502 (2012).

\bibitem{MDI_2018_2}
C. Lupo, C. Ottaviani, P. Papanastasiou, S. Pirandola,
Phys. Rev. Lett. {\bf 120}, 220505 (2018)

\bibitem{GaussDeFinetti}
A. Leverrier,
arXiv:1612.05080 (2016).

\bibitem{Lev2010}
A. Leverrier, F. Grosshans, P. Grangier,
Phys. Rev. A {\bf 81}, 062343 (2010).
 
\bibitem{Usenko}
L. Ruppert, V. C. Usenko, R. Filip,
Phys. Rev. A {\bf 90}, 062310 (2014).

\bibitem{NL1}
Y.-B. Zhao, M. Heid, J. Rigas, N. L\"utkenhaus,
Phys. Rev. A {\bf 79}, 012307 (2009).

\bibitem{Kamil}
K. Bradler, C. Weedbrook,
Phys. Rev. A {\bf 97}, 022310 (2018).

\bibitem{PP}
P. Papanastasiou, C. Lupo, C. Weedbrook, S. Pirandola,
Phys. Rev. A {\bf 98}, 012340 (2018).

\end{thebibliography}
\end{document}